# A Survey of Water Production in 61 Comets from SOHO/SWAN Observations of Hydrogen Lyman-alpha: Twenty-One Years 1996-2016

Short Title: Water Production in Comets 1996-2016

M.R. Combi[1], T.T. Mäkinen[2], J.-L. Bertaux[3], E. Quémerais[3], S. Ferron[4]

[1]Dept. of Climate and Space Sciences and Engineering

University of Michigan

2455 Hayward Street

Ann Arbor, MI 48109-2143

*Corresponding author: mcombi@umich.edu

[2]Finnish Meteorological Institute, Box 503

SF-00101 Helsinki, FINLAND

[3] LATMOS/IPSL

Université de Versailles Saint-Quentin

11, Boulevard d'Alembert, 78280, Guyancourt, FRANCE

[4]ACRI-ST, Sophia-Antipolis, FRANCE

6 Tables

9 Figures

**Abstract**

The Solar Wind Anisotropies (SWAN) instrument on the SOlar and Heliospheric Observatory (SOHO) satellite has observed 44 long period and new Oort cloud comets and 36 apparitions of 17 short period comets since its launch in December 1995. Water production rates have been determined from the over 3700 images producing a consistent set of activity variations over large parts of each comet's orbit. This has enabled the calculation of exponential power-law variations with heliocentric distance of these comets both before and after perihelion, as well as the absolute values of the water production rates. These various measures of overall water activity including pre- and post-perihelion exponents, absolute water production rates at 1 AU, active surface areas and their variations have been compared with a number of dynamical quantities for each comet including dynamical class, original semi-major axis, nucleus radius (when available), and compositional taxonomic class. Evidence for evolution of cometary nuclei is seen in both long-period and short-period comets.

**1. Introduction**

The Solar Wind ANisotropies (SWAN) instrument on board the SOlar and Heliospheric Observatory (SOHO) satellite has been operating in a halo orbit around the Earth-Sun L1 Lagrange point since shortly after its launch on 2 December 1995. Its primary science mission has been to provide continuous monitoring of the whole sky distribution of hydrogen Lyman-alpha emission resulting from interplanetary atomic hydrogen streaming through the solar system and being eaten away by charge exchange, electron impact and ionization by the sun and solar wind and being illuminated by the Sun's Ly-$\alpha$ emission. Because of the required sensitivity and the ability to observe the full sky, SWAN has also been an excellent platform from which to observe the hydrogen Ly-α comae of many comets since the beginning of 1996. After more than 20 years in operation it is still providing excellent measurements of hydrogen in both the interplanetary medium (Bertaux et al., 1995) and comets (Bertaux et al. 1998). SWAN has provided important observations of many individual comets. The SWAN observations allowed to determine that a total of $2.7 \pm 0.4\ 10^9$ kg of water ice was lost by comet 67P during one perihelion passage in average (Bertaux et al., 2015). Combined with the estimated area of 20 km$^2$, it yields an equivalent thickness of 15 cm of ice sublimated inside the nucleus and lost to space in one perihelion passage of this comet. SWAN observations also allowed for important coverage of the EPOXI mission target comet 103P/Hartley 2 from previous apparitions (Combi et al. 2011b) and providing coverage throughout the apparition of the flyby (Combi et al. 2011b). Finally, SWAN observations of comet 2012 S1 (ISON) by Combi et al. (2014) allowed for the determination of the water lost by the comet before its total disruption and loss at its very close perihelion passage, which is consistent with pre-perihelion estimates of the size of the nucleus (Lamy et al., 2014).

Most long-term surveys of cometary activity concentrate on composition (Newburn and Spinrad, 1989; A'Hearn et al. 1995; Fink and Hicks 1996; Fink 2009; Langland-Shula and Smith 2011; Cochran et al. 2012; Dello Russo et al. 2016), although A'Hearn et al. (1995) did examine heliocentric distance dependencies of water proxies like OH or even CN. The Nançay radio survey of 18-cm lines of OH in comets (Crovisier et al. 2002), on the other hand, covers OH observations and thus mainly water production rates. Similarly, this survey covers the water production rates of 61 comets each observed over extended periods of time allowing for exponential power-law exponents and production rates at 1 AU to be computed for pre- and post-perihelion legs of the comet orbits. Most of the long period comets included in this survey were observed after most of the comets included most of the published surveys so detailed comparisons with compositional taxonomic classes do not include many overlapping comets. One exception is the recent infrared survey of Dello Russo et al. (2016), however their IR taxonomy is rather involved providing depleted, typical and enhanced classification for 10 different molecular species compared to $H_2O$. Unfortunately even for the IR survey there are not enough common comets to make meaningful cross-comparisons. However, Dello Russo et al. looked at the literature for the visual observations of the common carbon radicals and list those taxonomic classes (depleted or typical) for several comets in the SWAN dataset.

**2. Observations and Basic Model Analysis**

The SWAN all-sky camera consists of two systems one for the north heliographic hemisphere and the other for the south. Each has a 5x5 array of detectors of one square degree each that are scanned across the sky every day yielding a full sky map of hydrogen Ly-$\alpha$ emission. Depending on location in the sky with respect to the Sun and the galactic equator, comets brighter than magnitude ~10-12 can be detected by SWAN. During the first 10 years of operation, SWAN was at times targeted specifically at comets with increased exposure time and spatial double-oversampling the one square degree pixels. The detector is actually a broadband far ultraviolet detector, which has sensitivity to either side of Ly$\alpha$, however for comets, nearly 98% of the total far ultraviolet emission is typically in Ly$\alpha$, and all the other emissions, e.g., O, C, S, CO, etc., are concentrated well within the innermost 1-degree pixel centered on the nucleus. For our analyses we normally sample the coma out to a radius of 8 degrees. The extra total non-Ly$\alpha$ signal is quite small and accounted for in our stated uncertainties. While comets are observed at quite a range of geocentric distance within this 8 degree radius aperture, this distance is nearly always larger than the effective production scale length of H atoms resulting from photodissociation of $H_2O$ and OH and nearly always smaller than the decay scale length of the outward streaming fast H atoms. Therefore, there should be no systematic error resulting from the sampling size of the production rate calculation. Such effects as they relate to typical photometric and spectrophotometric observations of the common visible radicals ($C_2$, $C_3$, CN, $NH_2$, NH and OH), were on the other hand shown to be quite sensitive to the choice of model parameters (Fink and Combi, 2004) and could yield systematic differences in determined abundances and production rates.

Over 90% of observed hydrogen atoms in comets are produced by the photodissociation of $H_2O$ molecules and the subsequent photodissociation of OH with typically 6%±3% coming from all other sources (Combi et al., 2005). Nascent H atoms are produced with a range of speeds of 8 - 20 km s$^{-1}$ resulting from the excess energy of photodissociation (Keller and Meier, 1976; Combi et al., 2004). The combination of photodissociation rate and water production rate, both of which increase as a comet's heliocentric distance decreases, determine the fraction of H atoms that collide with the heavy molecules, mostly water, in the inner coma. When this happens in sufficient numbers the transfer of energy results in the slowing of H atoms and heating of the heavy molecules, which results in a faster expanding coma. The shaping of the H atom speed distribution and the resulting physical effect on the expansion of the coma have been observed and quantified quite well (Bockelée-Morvan & Crovisier, 1987; Combi & Smyth, 1988a&b; Combi, 1989; Combi et al., 1998; Combi et al., 2000; Tseng et al., 2007; Shou et al., 2016).

SWAN images of the H Lyα coma are analyzed with a method called the time-resolved model (TRM) that was described in the paper by Mäkinen & Combi (2005) and combines aspects of several past modeling approaches, namely, the vectorial model (Festou, 1981); the syndyne model (Keller and Meier, 1976), and the particle kinetic physics models (Combi & Smyth 1988a&b). Input parameters to the model include the comet's orbital elements obtained from the JPL Horizons web site (http://ssd.jpl.nasa.gov/horizons.cgi), the SOHO orbit obtained from the SOHO project, and the daily solar Lyα flux at the Earth obtained from the LISRD web site at LASP, University of Colorado, (http://lasp.colorado.edu/lisird/lya/). The solar Lyα flux at the comet is estimated by taking the nearest value at the Earth facing the same part of the Sun correcting for the difference in heliographic longitude between the Earth and the comet. The fraction of H atoms thermalized by collisions with the water dominated coma and the resulting velocity distribution of H atoms leaving the inner coma is calculated from a parameterized version of the hybrid fluid/kinetic model calculations (Combi & Smyth, 1988a&b; Combi et al., 2000).

As an example Figure 1 shows the portion of the full-sky image of 1 November 2011 centered on comet C/2009 P1 (Garradd) from the results of Combi et al. (2013) that is part of the TRM analysis. The aperture for summing up the comet Lyα emission is an 8-degree radius circle centered on the comet, shown in gray. The outer region of the IPM background with field stars is shown in blue. The red areas are the locations of field stars that are masked off so as not to count those in the comet brightness for optimizing the model fit. The TRM fits a comet distribution as described above and a tilted background for the IPM in first or second order. Figure 2 shows the profile of brightness that corresponds to the cut shown as the red line in Figure 1. The observed comet profile is in white, the modeled comet profile in green and the subtracted IPM line shown as the darker straighter green line below. The masked star contribution is shown in red and regions outside the comet as part of the IPM are shown in blue. The result of each TRM model analysis is to produce a water production from each image.

In addition to calculating the average water production rate for each image the TRM, as described by Mäkinen and Combi (2005) also employs a method to analyze long sequences of images together to deconvolve the daily water production rates at the nucleus. When comets are bright enough this method can be used to track secular variations of the initial water production rate at the nucleus rather than just the average value responsible for the hydrogen distribution within several degrees of the nucleus. Results of this procedure have been published for a number of comets in the past, namely comet C/1996 B2 (Hyakutake) by Combi et al. (2005), comets 1999 H1 (Lee), 1999 T1 (McNaught-Hartley), C/2001 A2 (LINEAR), C/2000 WM1 (LINEAR) and 153P/Ikeya-Zhang by Combi et al. (2008) and C/2012 S1 (ISON) by Combi et al. (2014). In such cases we can compare with other observations taken of shorter-lived species and closer to the nucleus and compare shorter-term temporal variations such as outbursts. Making such comparisons shows that the normal single image production rates delays and averages the water production rate over a time period of about 1 to 3 days depending on the distance to the comet. Therefore, an abrupt outburst is both delayed and smoothed out over up to 3 days as seen in the response of a single hydrogen coma image. None of the results given in this survey depend on the deconvolution processing.

The remainder of this paper describes summary results of 44 new/long-period comets and 39 apparitions of 17 short period comets observed by SOHO/SWAN since 1996 and analyzed with the TRM. Most of the data have already been archived and certified in the Small Bodies Node (SBN) of the NASA Planetary Data System (PDS), which can be found at Combi (2017). The results of 9 newer long-period comets observed since 2012 have been recently published by Combi et al. (2018) and will be submitted to the PDS in 2018. Results of the most recent apparitions of 4 short period comets (2P, 45P and 96P) are included in this summary and will be the subjects of a future paper and will then be submitted to the PDS. In all three cases the most recent apparitions of these comets were quite similar to the previous ones.

When possible pre- and post-perihelion variations in water production rate have been fitted with a power-law of the form $Q = Q_1 r^p$., where $Q_1$ is the value at a heliocentric distance of 1 AU, r is the heliocentric distance in AU and p is the power-law exponent, to which we often refer as "slope," but this slope is not directly related directly to slopes in visual light curves. Some comets never reached a heliocentric distance of 1 AU and were either always closer or farther away, so the comet never actually had a production rate at 1 AU. In these cases the 1 AU extrapolated value is given anyway. In a few cases such a fit was not possible for a number of reasons. In some cases observational geometry prevented enough measurements to be made either before or after perihelion. In other cases the variation was too irregular for a power-law to have any physical meaning. An example of this would be a comet where the seasonal effects dominate the variation over time. In other cases the range of heliocentric distance obtained was not large enough for a meaningful power law to be obtained. For 3 long-period comets we provide a mean value at 1 AU for pre- and/or post-perihelion but not a power-law for the reasons stated above. These are C/2001 OG108 (LONEOS), C/2002 G3 (SOHO), and C/2002 O4 (Hoenig).

The error bars for the fitted slopes are given in the appropriate tables. Formal error bars for the fitted production rates are all quite small ranging between 1% and 5%. These are comparable to the individual uncertainties for each production rate given in the PDS tables of individual images and are indicative of internal systematic error resulting from the model fitting, background subtraction procedure and stochastic noise in the SWAN brightness. These are also indicative of the relative uncertainties comparing different SWAN water production rates with one another. On the other hand absolute values of active areas have uncertainties as large as the uncertainties in our production rates, owing to uncertainties in model parameters such as molecular lifetimes, the intensity calibration of SWAN, uncertainty in the adopted absolute value of the solar flux from the LASP web site that contributes to the Lyman-alpha fluorescence used and the 6±3% uncertainty (Combi et al., 2005) coming from other sources of H atoms. Altogether we normally give an approximate absolute uncertainty to water production rates as ±30%. This would be indicative of comparisons of our water production rates with those determined by other observations, such as ground-based IR observations of $H_2O$, ground-based or space-based observations of OH or radio observations of OH. Because of the large ranges of values in the figures, most of the error bars are comparable to or smaller than the sizes of the data points.

For some short period comets of multiple apparitions it is clear that there is not much overall change from apparition to apparition, e.g. 2P, 21P, 45P, 46P and 96P. Data from multiple apparitions were taken together for calculating power-law variations with heliocentric distance. In section 4 below these are compared with comets 19P, 55P and 141P that were only observed for one apparition each for purposes of correlation analyses.

For dynamical classification we have adopted the criteria given in the photometric survey study of A'Hearn et al. (1995) that is based on the value of the original semi-major axis ($a_0$) of each comet. These values were determined from two sources: the JPL Horizons web site and the Minor Planet Center web site. A'Hearn et al. (1995) used values provided by Brian Marsden of the Minor Planet Center. The Minor Planet Center (MPC) provides values of $a_0$ for most of the comets. With the JPL Horizons tools one can calculate the osculating orbital elements at any desired date. For this work we picked a date of 1950.0 where all the comets in the survey were well past 50 AU on their inbound orbits and long before any influence of Neptune. In most cases the numerical values of $a_0$ from the two sources were well within 10% of one another. For these we simply give the average of the JPL and MPC values. In a few cases $a_0$ was very large, or more appropriately the calculated $1/a_0$ was very close to zero or even slightly negative (hyperbolic). It is expected that most hyperbolic orbits resulted from a stellar or planetary perturbations and not indicating original extra-solar comets. However, such comets would eventually leave the solar system. In a couple of cases either there was no MPC value so we give only the JPL Horizons value. When there was some difference in the numerical value of $a_0$ between the two sources, the dynamical class was still the same.

The A'Hearn et al. dynamical classes are defined as follows:

DN - dynamically new: $a_0 > 20000$ AU; $1/a_0 < 50 \times 10^{-6}$

YL - young, long period: $20000$ AU $> a_0 > 500$ AU; $50 \times 10^{-6} < 1/a_0 < 2000 \times 10^{-6}$

OL – old, long period: $a_0 < 500$ AU; $1/a_0 < 2000 \times 10^{-6}$

JF - Jupiter family $2.0 < T_J < 3.0$

HF - Halley family $T_J > 2.0$

where $a_0$ is the original semi-major axis for long period comets before entering the planet region of the solar system, $T_J = \frac{a_J}{a} + 2\sqrt{\frac{a}{a_J}(1-e^2)} \cos i$, is the Tisserand constant with Jupiter, $a_J$ is the semi-major axis of Jupiter, a, e and i are the semi-major axis, eccentricity and inclination of the comet orbit, respectively. While the dynamically new class is based on a large original semi-major axis, A'Hearn et al. (1995) point out that this only means there is a 90% chance of a comet so classified is really on its first trip into the inner solar system. By the same token other long period comets, YL or OL, could have been only through the outer solar system previously and had their orbits changed to a smaller values of $a_0$ but did not have a small enough perihelion distance to receive significant solar processing. Therefore, all dynamical classifications are necessarily probabilistic and not an exact indication of past thermal evolution.

Similarly, it is important to stress that the pre- and post-perihelion power-law results here also need to be understood only as ensemble properties of various populations of the comets observed and are indicative of a wide variety of physical responsible for cometary activity. For any individual comet, the pre- and post-perihelion power-laws could be heavily influenced by peculiar geometrical factors such as the angle between the spin axis and the orbital axis yielding seasonal effects owing to solar illumination patterns. Comet 17P/Holmes, for example, was only observed in 2007 in the aftermath of a huge outburst (Combi et al. 2007) and so a power-law variation with heliocentric distance makes little sense. Results from the Rosetta mission to comet 67P/Churyumov-Gerasimenko have shown that nucleus activity and its variation with heliocentric distance might very well be influenced by diurnal variations as the solar heat wave penetrates to different depths and by moderate scale changes in the surface due to refreshing of more icy vertical surfaces after mass wasting falls.

De Sanctis et al. (2015) have shown the deposition of surface and/or near surface ice from infrared spectra obtained with the Rosetta VIRTIS spectrometer as regions of the comet rotate into darkness but at depth warmer areas continue to sublimate leaving more water closer to the surface. When the region returns into sunlight on the next rotation the water signature decreases in an hour indicating that the water then sublimates away. Similarly, but on a much larger scale, Fornasier et al. (2016) have shown from multicolor imaging of 67P with the Rosetta OSIRIS camera that the nucleus is overall "bluer" near perihelion than farther from perihelion indicating that the surface layers are enriched with water ice near perihelion when the heat wave penetrates much deeper. This might explain why the overall heliocentric distance variation of 67P is so steep, having a slope of -4.5 (Fougere et al. 2016a and 2016b).

Images of the surface of 67P obtained by the Rosetta OSIRIS camera show the continuous evolution of a complex surface with evidence of activity on steeper vertical cliffs and the resulting material falls at the bottom of the cliffs. Vincent et al. (2016) have suggested that cometary dust jets actually result from activity on vertical fractured cliffs as well as in depressions because material can fall rather than leaving a covering dust mantle that could insulate those regions on the surface. Based on the increased solar insolation exposure of the southern hemisphere the difference in steep slope structures between the northern and southern hemispheres have been quantified by Vincent et al. (2017). Even at larger heliocentric distance evidence for similar surface reprocessing in the form of ice and rock flow and fall has been very recently reported by Raponi et al. (2018) at the surface of Ceres by the Dawn spacecraft.

## 3. SOHO/SWAN Survey Results – Long Period Comets

Table 1 contains the observational circumstances, dynamical classes and water production rate power-law results for 42 long-period comets observed by SWAN and analyzed over the past 20 years. The dataset is based on over 2600 images. Several comets were observed for a year or more with over 100 observations each, e.g. C/1995 O1 (Hale-Bopp), C/2001 Q4 (NEAT), C/2009 P1 (Garradd), C/2012 Q2 (Lovejoy), etc. Some others are represented by only a handful of images.

Figure 3 shows the pre-perihelion slopes as a function of the inverse original semi-major axes for the entire set of long-period comets. Figure 4 shows the post-perihelion slopes as a function of the inverse original semi-major axes for the entire set of long-period comets. In both cases comets with very large semi-major axis and conversely very small inverse semi-major axis are all plotted just to the right of the plot at $a_0$ = 60000 AU, because the uncertainties in inverse semi-major axis are large. Figure 3 clearly shows a difference in pre-perihelion power-law slopes as a function of dynamical classes. The broadest variation in slopes ranging to values as low as -8 is confined to the OL class and being consistent with the most evolution. For those in the DN class slopes typically are in the range of -3 to -1. Those in the YL class are generally intermediate in range of slopes between the OL and DN classes. The behavior is consistent with a gradual steepening of the water production rate variation slope with age.

Figure 5 shows a scatter-correlation plot of pre-perihelion slope versus post-perihelion slope. There is no obvious correlation between the two. The main difference is that except for comet C/2014 Q1 (PanSTARRS) with both slopes being very steep at -7.8 and -8.9, the majority of pre-perihelion slopes cover a somewhat narrower range than do the post-perihelion slopes. Note that the lower limit of the post-perihelion slope does not include comet C/1999 S4 (LINEAR) at -19.6 resulting from its total disintegration beginning very near perihelion. It's very flat pre-perihelion slope of -1.2 as well as its very large original semi-major axis are consistent with it having been truly dynamically new and on its first trip into the inner solar system.

**4. SOHO/SWAN Survey Results – Short Period Comets**

Table 2 gives the observational and orbital aspects and parameters of each of the short period comet apparitions that were observed by the SOHO SWAN instrument since launch in December 1995. As can be seen from the value of the Tisserand constant with Jupiter and the classification definition given in the previous section of this paper, all of the comets are Jupiter Family with the exceptions of 55P/Tempel-Tuttle, 8P/Tuttle and 96P/Machholz 1. All these are well-known exceptions with 8P/Tuttle and 55P/Tempel-Tuttle being Halley Family comets and 96PMachholz 1 being a possibly captured Oort cloud or even extra-solar comet (Schleicher 2008). Unlike some previous comet survey papers, we did not include population histograms of quantities such as inclination, Tisserand constant, or perihelion, because the results are not particularly illuminating.

Table 3 gives observational results of the individual apparitions of short period comets including the power-law fits to pre- and post-perihelion parts of the orbit when available and appropriate as well the pre- and post-perihelion active areas from the production rate at 1 AU using the method of Cowan and A'Hearn (1979) for a rotating nucleus. Table 4 gives a reduced set of short period comet results where the power-law fits for five multiple apparition comets have been taken together as well as where nucleus radii have been determined. We did not include comets 41P, 46P, or 73P, which show considerable variation from apparition to apparition.

The trends of pre- or post-perihelion power law slopes with the short period comets' perihelion distance are weak. One might expect if increasing slope were a sign of evolution that the comets with smaller perihelion distances might tend to have steeper slopes, but the perihelion distance and slope are not correlated. In Figure 6 there is no obvious trend of pre-perihelion power-law slope with the perihelion distance. In Figure 7 there is a slight downward trend of post-perihelion power-law slope with the perihelion distance, but it is difficult to conclude it is significant. The short period comets in Figures 6 and 7 include multiple apparition comets (Table 4), which seem consistent over the apparitions covered, and single apparition comets, where there was a reasonable power-law fit.

The same cannot be said about the active fraction, however. For the short period comets where there are measured values of the radius, it is possible to calculate the active fractional area of the comet by using the Cowan and A'Hearn (1979) method for calculation of the active area assuming a rapidly rotating nucleus and the power-law fitted production rate at 1 AU, and dividing by $4\pi r_N^2$, where $r_N$ is the radius of the comet, and water-driven sublimation from a dark nucleus. Errors in active fraction are driven by uncertainties in water production rates, determined nucleus radii, and the appropriateness of the Cowan and A'Hearn (1979) sublimation models used. Published nucleus radii are determined by one or more of several methods (Lamy et al. 2004): photometry of bare nuclei at large heliocentric distance, combinations of visual photometry and thermal infrared flux, radar cross sections, and spacecraft flyby imaging.

Figures 8 and 9 show plots of the active fraction plotted against each comet's perihelion distance for the pre-perihelion and post-perihelion power-laws, respectively. For the comets that appear in Figures 8 and 9 we include error estimates in the notes to Table 3 that come from the original published papers or the mean of several published measured values. Comets 8P and 41P have radar cross sections. Radii of comets 19P, 67P and 103P have been determined by several methods but were verified in spacecraft flyby images. Radii of comets 2P and 21P were determined by traditional ground-based methods but have quite small uncertainties of ± 0.3 and 0.05 km, respectively. The radius of comet 96P was also determined by traditional ground-based optical methods but Lamy et al. (2004) gives three very similar values of 3.5, 2.8 and 3.2 km which yield a mean error of ± 0.2 km. Similarly, 46P which was the original Rosetta mission target comet has values of 0.62, 0.56, 0.7 and 0.6 km as well as some much larger upper limits. The average of 0.6 km has a mean uncertainty of ± 0.04 km. The two most uncertain are comet 41P with an optical value of 0.7 km and a lower limit of > 0.45 km from radar by Howell et al. (2017). Therefore for most of the comets in Figure 8 and 9 the uncertainties in radius are in the range of up to 20% which would yield an additional uncertainty to the active fraction of up to ~40%. Since the range of abscissae in Figures 8 and 9 cover three orders of magnitude the apparent trends are quite valid even accounting for all sources of uncertainty. For both pre- and post-perihelion cases there are clear and obvious trends of small active fractions for small perihelion distances. This could very well be consistent with the increased solar exposure of comets with smaller perihelion distances on each orbit gradually decreasing the near surface coverage of sublimating ice and thus lowering activity levels per unit surface area.

5. **Compositional Taxonomic Classes**

Table 5 gives the standard ground-based carbon species taxonomic classifications of the set of comet observed by SWAN, namely typical and depleted. For this we have combined classifications found in the papers by A'Hearn et al. (1995), Fink and Hicks (1996), Fink (2009), Cochran et al. (2012), and a recent reclassification of visual ground based results from various published data by Dello Russo et al. (2016). The comprehensive multi-level classification of parent molecules in the infrared by Dello Russo et al. (2016) does not have enough comets in common with the SWAN set to make a meaningful comparison. Classification for C/2009 R1 (McNaught) was taken from the observations of Korsun et al. (2012) and that for C/2014 Q2 (Lovejoy) was taken from the observations of Venkataramani et al. (2016). Table 5 shows the SWAN comets for which typical and depleted classifications can be inferred by measurements of $C_2$, CN and OH.

Table 6 shows pre- and post-perihelion slopes comparing short period, typical and depleted comets and long period, typical and depleted comets. Unfortunately, the averages are limited by small number statistics. There are some trends, however. First, in each taxonomic class there is no significant difference between pre- and post-perihelion populations of each, so in Table 6 we averaged pre- and post-perihelion slopes to increase the sampling set. Perhaps surprisingly for long-period comets the depleted comets had shallower average power-law slopes than did the typical comets. The depleted short period comets had steeper slopes on the average than the typical short period comets. In additions, not surprisingly, the short period comets had steeper average power-law slopes than long period comets in general. For the comparison of long and short period comets the same trend was seen in the photometric survey of A'Hearn et al. (1995).

**6. Summary and Conclusions**

The photometric composition survey of A'Hearn et al. (1995) provided a comparison of a limited number of production rate power-law exponent dependencies for OH, CN $C_2$, $C_3$ and NH, where OH can be considered a reasonable proxy for $H_2O$. For OH they found a trend of steeper negative slopes going from Dynamically New pre-perihelion, Dynamically New post-perihelion Young Long-Period, Old Long-Period, Halley Family, Jupiter Family typical, and Jupiter Family depleted. In these cases their power-law exponents were generally only determined from a few points at different heliocentric distance. Our results shown in Figures 8 and 9 are consistent with the trend of the median values shown by A'Hearn et al. but having many more samples determined from wider and higher sampled production rate values shows the details are more complicated. We show a much wider range of slopes for long period comets with smaller original semi-major axes that would be classified as Old Long Period and a narrow range for those with large original semi-major axes that would be classified as Dynamically New or Young Long Period. We do see an average difference between pre- and post-perihelion dynamically new comets, but having more samples shows there is still a considerable range in values about the mean or median.

For the short-period comets where nucleus radii have been determined there is a clear and consistent trend of larger active fractional area with larger perihelion distance. This would be consistent with an evolutionary trend of more processing for the comets with smaller perihelion distance caused by the larger orbit-integrated solar insolation. There is, however, little correlation of power-law slope with perihelion distance. For the long-period comets the main evolution trend is that the Old Long Period comets have a much wider range of power-law slopes than to Dynamically New or Young Long Period comets.

A possible difference in the manifestation of the evolutionary changes between short-period and long-period comets could be at what stage the evolution happened. The main effect of solar exposure on long-period comets we see likely happens on the comets' first or first couple of passes into the inner solar system. This is when the primordial surface layers that have long been at large distances from the sun are first sublimated away. In SWAN survey results there is both a small but significant difference in the pre- and post-perihelion power-law slopes as well as most clearly a change in the range of power-law slopes in progressing from the Dynamically New to Young Long Period and finally to Old Long Period. Because of their short periods most short period comets had their first pass or two in the inner solar system at least several passes ago and fell into whatever their slopes are based mostly on seasonal effects and distribution of active regions on the surface. The evolutionary trend of smaller active fractional areas with smaller perihelion distances is possibly seen after many more orbits around the sun. The original more primordial surface layers of short period comets coming from the Kuiper belt are likely burned off after the first pass through the inner solar system upon capture by Jupiter.

Comparisons of power-law slopes with compositional taxonomic classes are complicated by the more limited overlap in comets observed by SWAN and those that have been classified by compositional surveys especially for the long-period comets. Even though there was difference between the distribution of pre- and post-perihelion slopes for the whole long-period comet set, for those which have a taxonomic classification there was no significant difference in pre- and post-perihelion slopes, so both were averaged together for the purpose of comparison here. In this case the depleted long-period comets had shallower slopes than the typical long-period comets. This could be simply a difference with no evolutionary cause. For the short period comets that have been classified by composition, the comparison was the opposite where the depleted short-period comets had steeper slopes than the typical short-period comets. It has been suspected that chemical depletion of volatile species may have no consistent evolutionary trend or show a difference between Oort cloud (long period) and Kuiper Belt (short period) origin (Mumma and Charnley 2011; Dello Russo et al. 2016). There is nothing in the results contained in the SWAN survey that goes against this idea; however there do seem to be consistent trends in power-law slopes and active fraction that are related to the evolution of the cometary nucleus.

**Acknowledgements**

SOHO is a cooperative international mission of ESA and NASA. M. Combi acknowledges support from grant NNX15AJ81G from the Solar System Observations Planetary Astronomy Program and NNX13AQ66G from the Planetary Mission Data Analysis Program. T.T. Mäkinen was supported by the Finnish Meteorological Institute (FMI). J.-L. Bertaux and E. Quémerais acknowledge support from CNRS and CNES. We obtained cometary ephemerides, orbital elements and original semi-major axes from the JPL Horizons web site (http://ssd.jpl.nasa.gov/horizons.cgi). Cometary original semi-major axes were also obtained at the IAU Minor Planet Center Database Search (https://www.minorplanetcenter.net/db_search). The composite solar Lyman-alpha data was taken from the LASP web site at the University of Colorado (http://lasp.colorado.edu/lisird/lya/). We also acknowledge the personnel that have been keeping SOHO and SWAN operational for over 20 years, in particular Dr. Walter Schmidt at FMI.

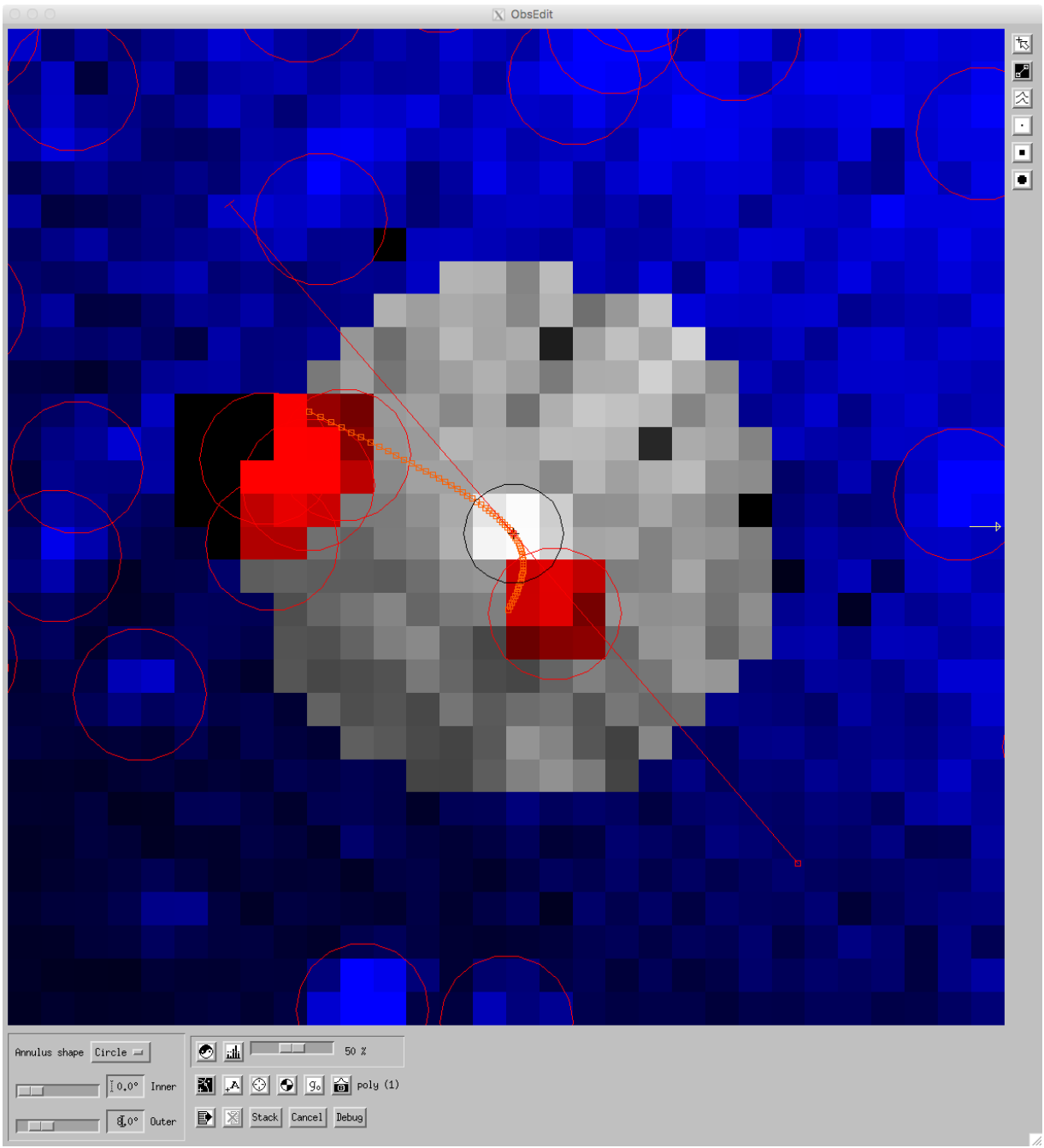

Figure 1. Lyman-α brightness distribution around comet C/2009 P1 (Garradd) on 1 November 2011. The comet is at the center of the image. The analysis is done on circle of radius 8 degrees shown in gray shades. Each small square corresponds to a 1 degree pixel. The blue shades are the sky background, namely the interplanetary medium emission of hydrogen Lyα and background stars. The locations of stars are indicated by red circles. Those stars that would interfere with the comet signal are masked and highlighted in red. The red line from upper left to lower right shows the location of the brightness profile shown in Figure 2.

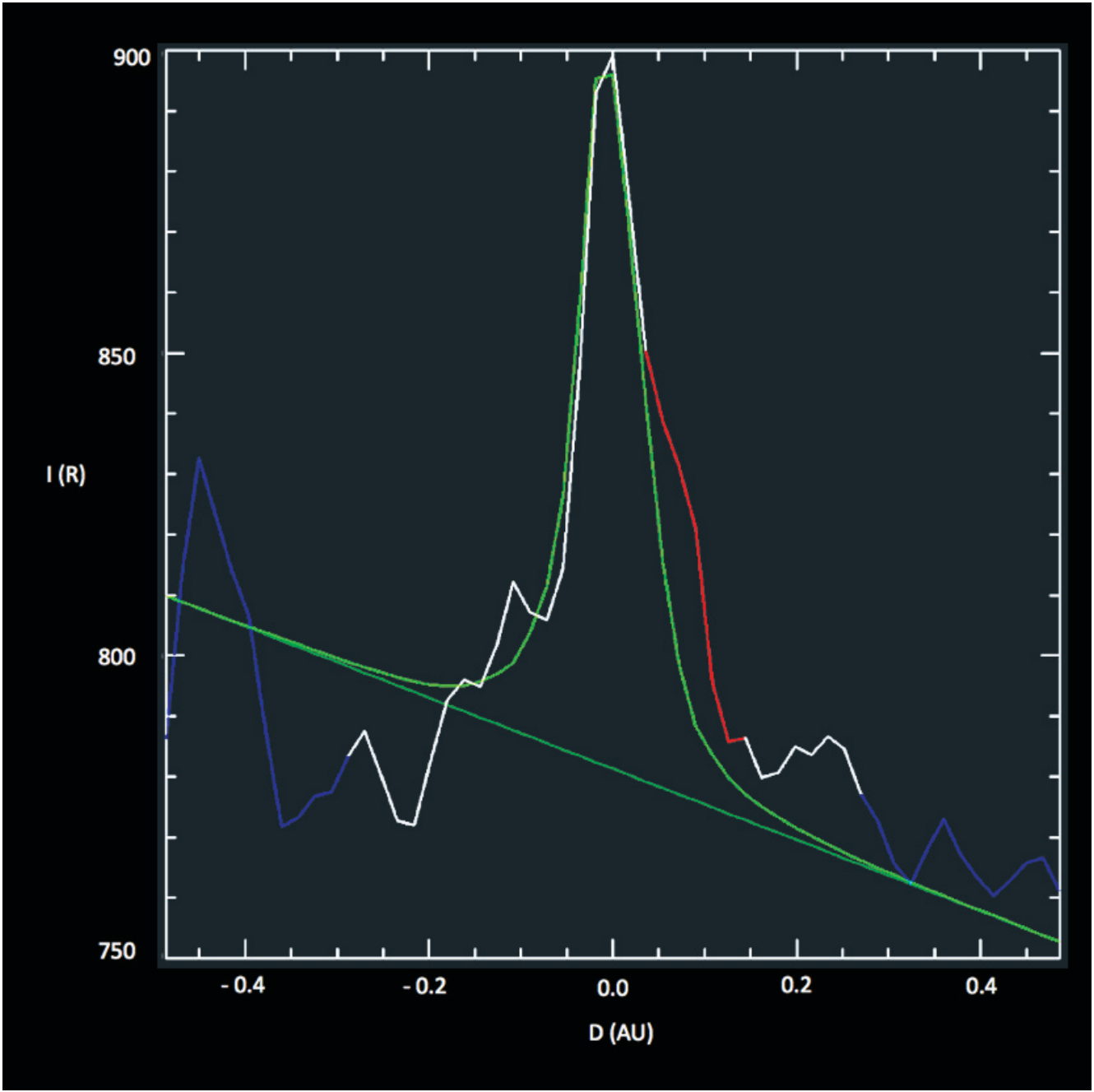

Figure 2. A brightness profile through the position of comet C/2009 P1 (Garradd). The white line indicates the observed brightness distribution. The green peaked profile is the TRM fit to the comet distribution and the green line below is the TRM fit to the IPM background. Red lines indicate the presence of stars that are excluded from the analysis. The blue line is the observed brightness mainly of the background IPM. Brightness is in Rayleighs and distance is in AU.

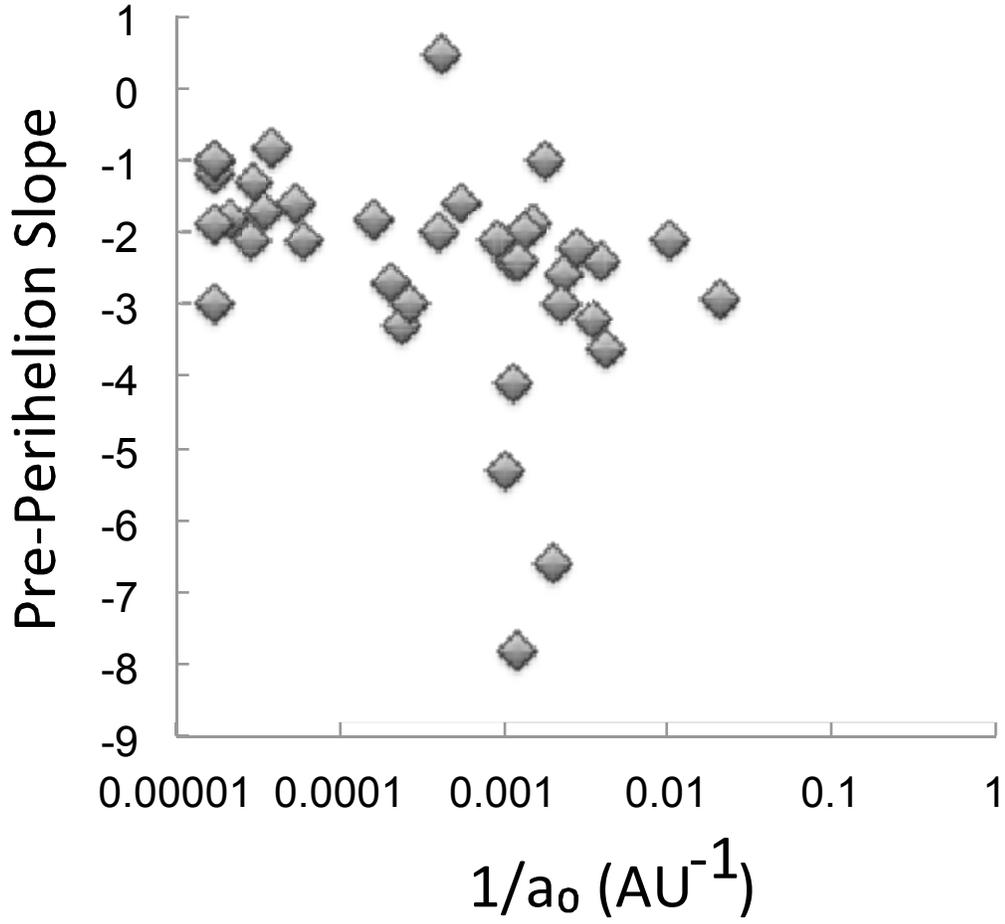

Figure 3. Dependence of the variation of pre-perihelion slopes on long-period comet inverse original semi-major axis.

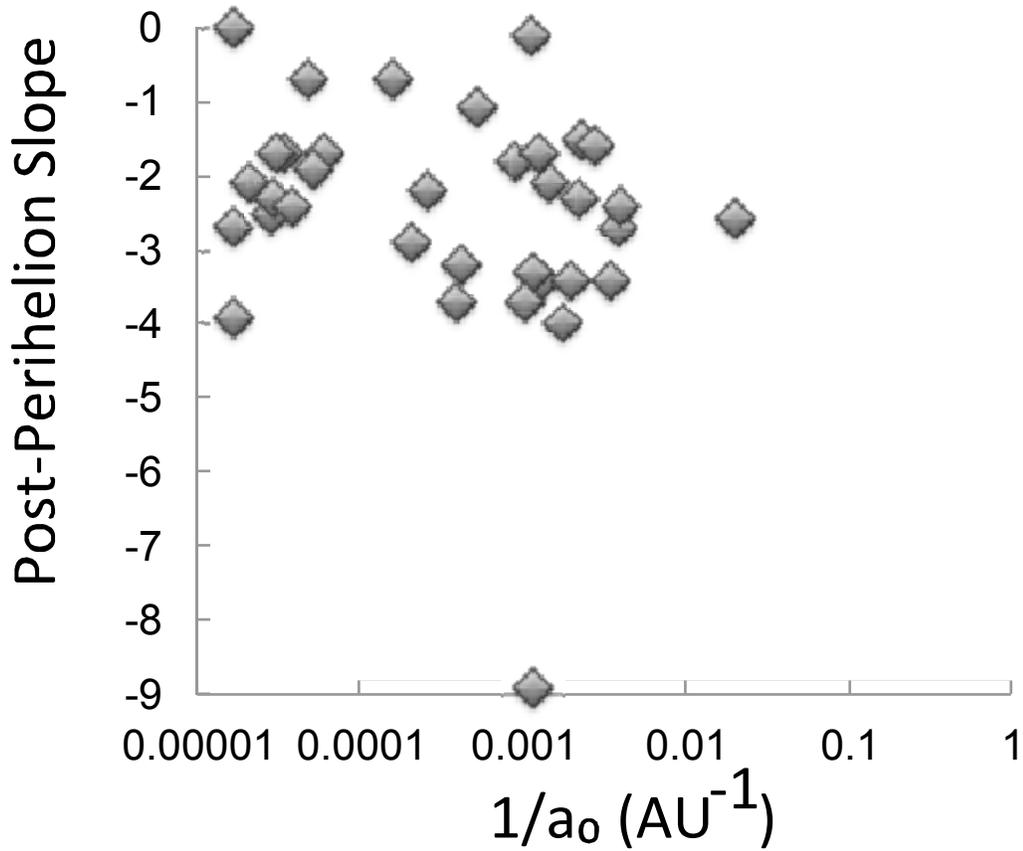

Figure 4. Dependence of the variation of post-perihelion slopes on long-period comet inverse original semi-major axis. The value of the post-perihelion slope for comet C/1999 S4 (LINEAR) is -19.6 and does not appear because it is a special case and for easier direct comparison between Figures 3 and 4.

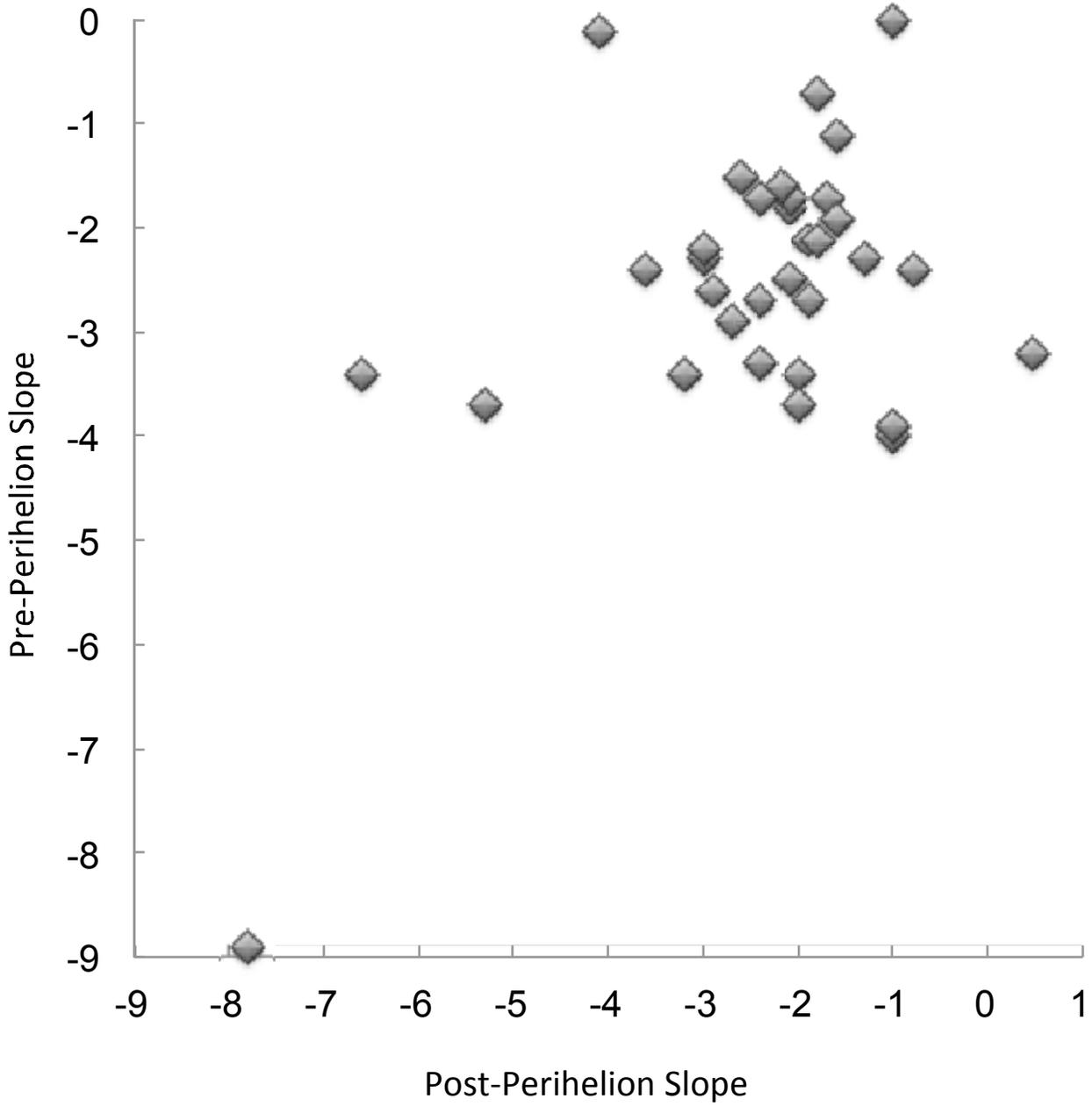

Figure 5. Correlation of pre-perihelion slopes with post-perihelion slopes for long-period comets.

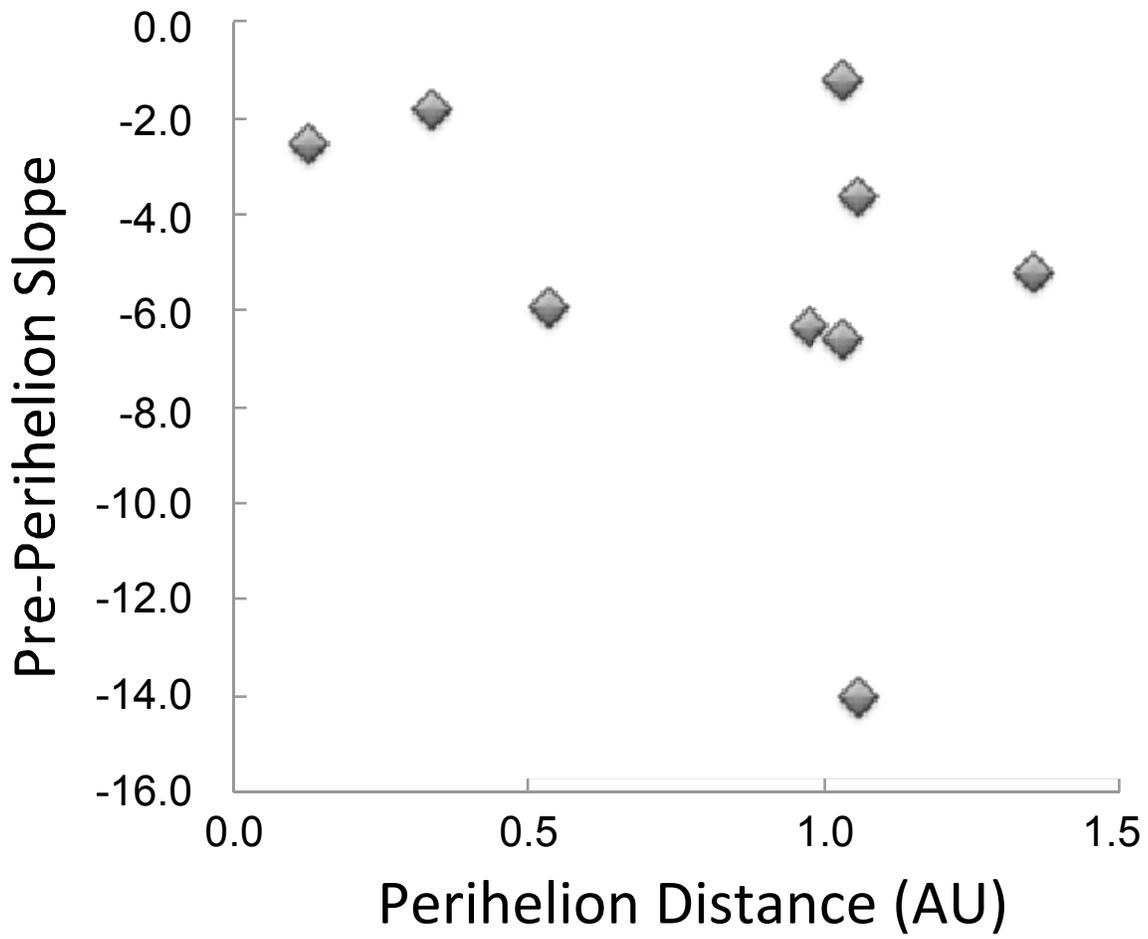

Figure 6. Correlation of pre-perihelion slopes with perihelion distance for short-period comets.

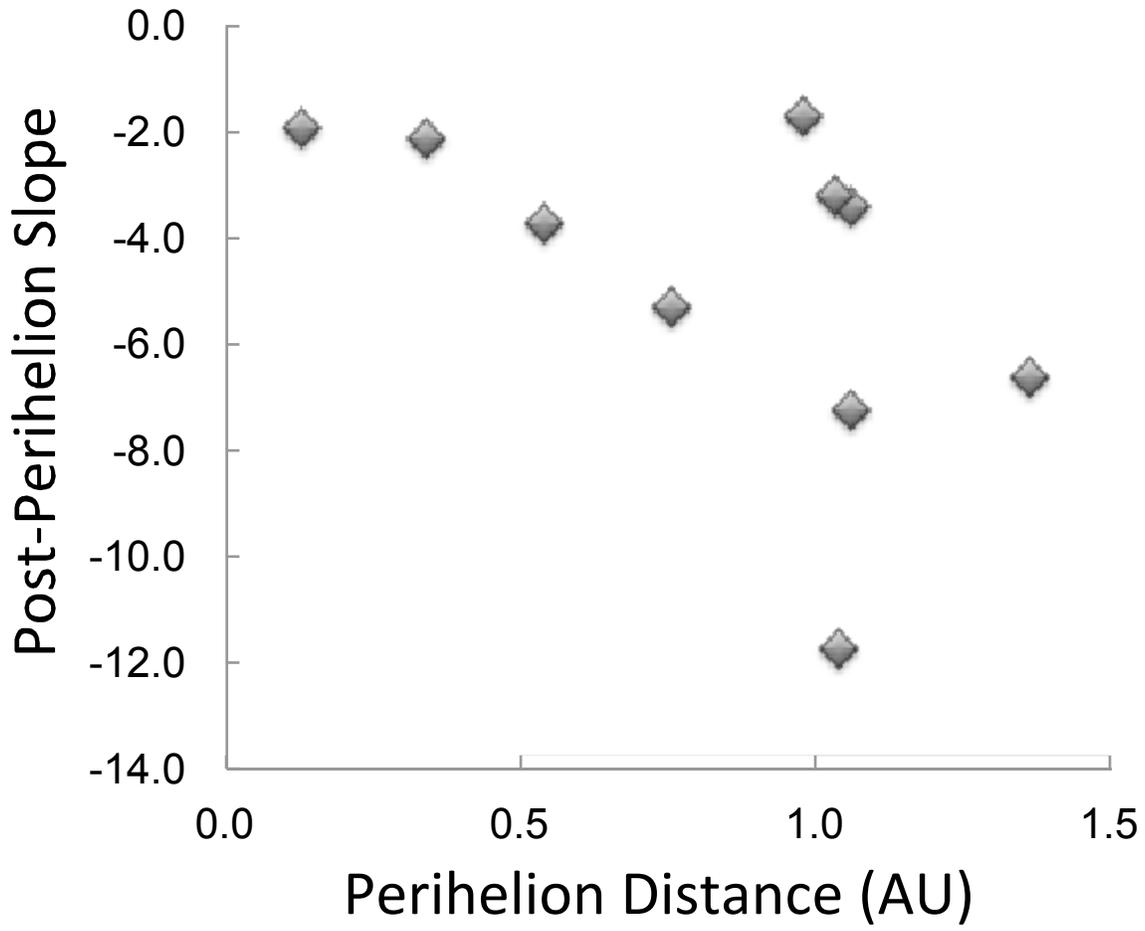

Figure 7. Correlation of post-perihelion slopes with perihelion distance for short-period comets.

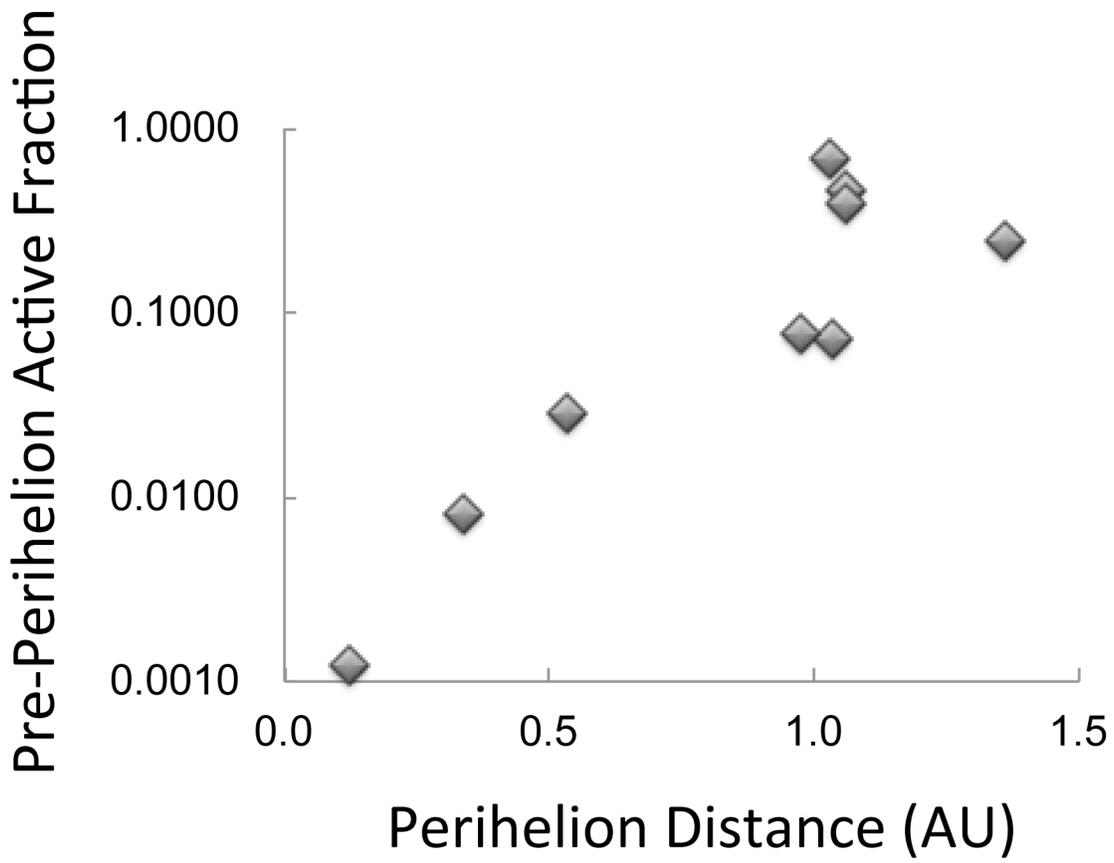

Figure 8. Correlation of pre-perihelion active fraction with perihelion distance for short-period comets.

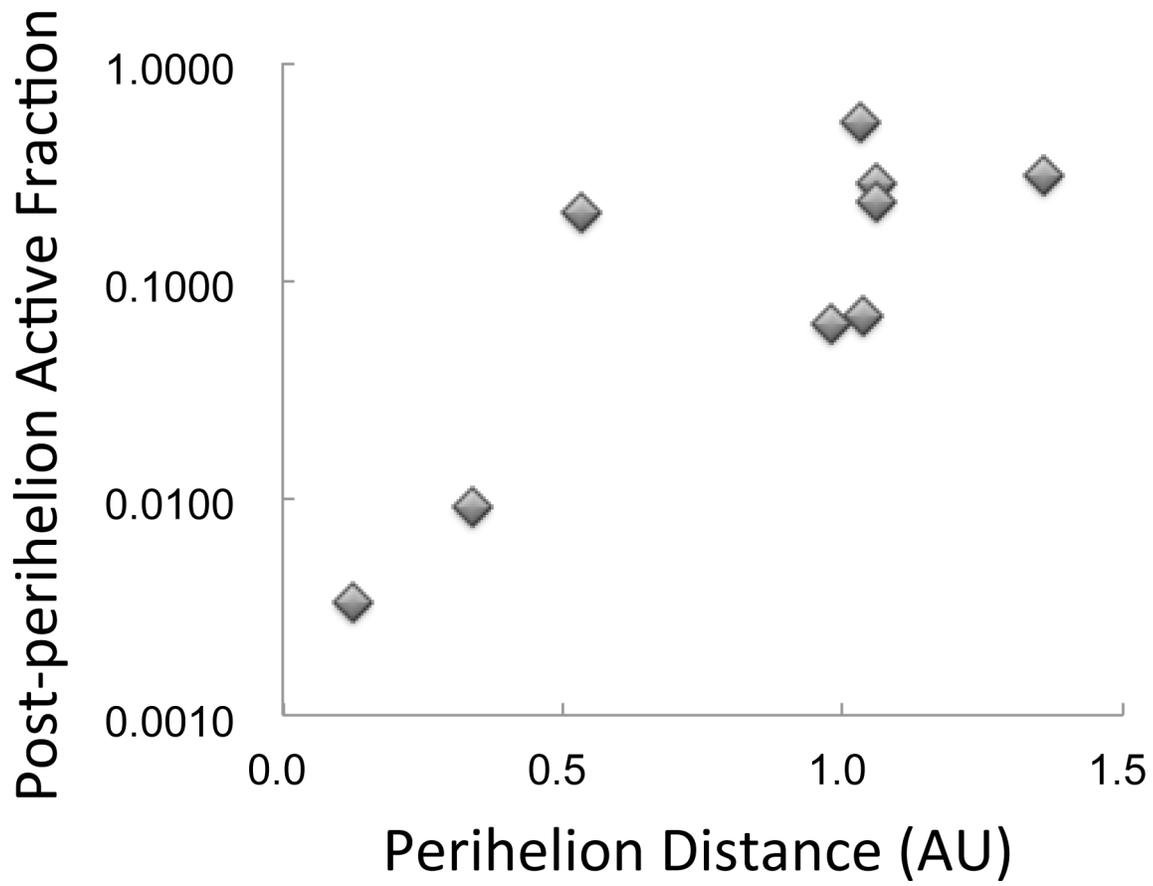

Figure 9. Correlation of post-perihelion active fraction with perihelion distance for short-period comets.

Table 1. Summary of Long-period Comets Observations, Orbital Aspects, and Production Rates

| Comet | T-Perihelion | T-Begin | T-End | N | q | $1/a_0$ | Class | Q1-pre | p-pre | Q1-post | p-post |
|---|---|---|---|---|---|---|---|---|---|---|---|
| C/1996 B2 (Hyakutake) | 19960501.3948 | -75.60 | 66.58 | 63 | 0.230224 | 1.52E-03 | OL | 5.05E+29 | -1.9±0.08 | 4.12E+29 | -2.1±0.08 |
| C/1996 Q1 (Tabur) | 19961103.5291 | -59.535 | 11.373 | 32 | 0.839815 | 1.80E-03 | OL | 4.13E+28 | -1.0±0.1 | 5.63E+27 | -4.0±2.2 |
| C/1995 O1 (Hale-Bopp) | 19970331.2270 | -240.124 | 178.316 | 142 | 0.922110 | 3.82E-03 | OL | 1.41E+31 | -2.4±0.02 | 1.24E+31 | -2.7±0.1 |
| C/1997 O1 (Tilbrook) | 19970713.4193 | -14.095 | 30.458 | 16 | 1.371803 | 1.14E-02 | OL | 7.30E+28 | a | 8.10E+28 | a |
| C/1998 U5 (LINEAR) | 19981221.7601 | -31.428 | -4.076 | 11 | 1.236452 | 1.01E-02 | OL | 4.75E+28 | -2.1±0.2 | a | a |
| C/1999 H1 (Lee) | 19990711.1732 | -66.544 | 69.933 | 50 | 0.708073 | 1.31E-03 | YL | 2.86E+29 | -2.0±0.2 | 2.35E+29 | -3.4±0.1 |
| C/1999 N2 (Lynn) | 19990723.0505 | -44.041 | 65.450 | 44 | 0.761284 | 3.45E-03 | OL | 4.11E+28 | -3.2±0.4 | 3.72E+28 | -3.1±0.2 |
| C/1999 J3 (LINEAR) | 19990920.1662 | -36.386 | 61.730 | 28 | 0.976811 | 1.15E-03 | YL | 3.26E+28 | -4.1±0.5 | 3.56E+28 | -0.13±0.3 |
| C/1999 S4 (LINEAR) | 20000726.1685 | -66.239 | 19.448 | 48 | 0.765007 | 3.21E-06 | DN | 3.22E+28 | -1.2±0.2 | 4.02E+25 | -19.6±5.2 |
| C/1999 T1 McNaught-Hartley | 20001213.4205 | -40.428 | 62.792 | 52 | 1.173711 | 1.15E-03 | YL | 5.76E+29 | -2.4±0.7 | 5.28E+29 | -3.3±0.2 |
| C/2001 A2 (LINEAR) | 20010524.5240 | -61.000 | 64.906 | 67 | 0.778609 | 1.07E-03 | YL | 7.35E+28 | -5.3±0.4 | 1.44E+29 | -3.7±0.1 |
| C/2000 WM1 (LINEAR) | 20020122.6773 | -51.696 | 46.379 | 60 | 0.555310 | 5.32E-04 | YL | 1.08E+29 | -1.6±0.1 | 2.36E+29 | -1.1±0.2 |
| C/2001 OG108 (LONEOS) | 20020315.0978 | -42.445 | 28.197 | 11 | 0.994847 | 7.65E-02 | OL | 2.35E+28 | a | 6.75E+27 | a |
| C/2002 C1 IP153) Ikeya-Zhang | 20020318.9828 | -44.542 | 120.498 | 63 | 0.507934 | 1.97E-02 | OL | 1.58E+29 | -2.9±0.2 | 2.63E+29 | -2.6±0.1 |
| C/2002 G3 (SOHO) | 20020416.7000 | -40.270 | -9.068 | 15 | 0.079400 | 1.02E-03 | YL | 2.18E+29 | a | b | a |
| C/2002 O4 (Hoenig) | 20021001.9400 | -72.876 | 17.858 | 28 | 0.775735 | -7.64E-04 | DN | 1.58E+28 | a | 5.30E+29 | a |
| C/2002 X5 (Kudo-Fujikawa) | 20030128.9761 | -41.691 | 50.615 | 34 | 0.190286 | 1.07E-03 | YL | 9.19E+28 | -2.1±0.2 | 5.53E+28 | -1.8±0.2 |
| C/2002 V1 (NEAT) | 20030218.2881 | -47.029 | 65.061 | 38 | 0.099256 | 2.29E-03 | OL | 8.62E+28 | -2.6±0.1 | 1.78E+29 | -1.5±0.2 |
| C/2002 Y1 (Juels-Holvorcem) | 20030413.2491 | -88.291 | 94.668 | 62 | 0.713667 | 4.11E-03 | OL | 9.57E+28 | -3.6±0.1 | 1.95E+29 | -2.4±0.1 |
| C/2002 T7 (LINEAR) | 20040423.0763 | -141.379 | 105.474 | 68 | 0.614585 | 2.89E-05 | DN | 1.10E+30 | -1.8±0.1 | 3.55E+29 | -2.1±0.1 |
| C/2001 Q4 (NEAT) | 20040515.9546 | -212.872 | 170.986 | 147 | 0.961886 | 6.04E-05 | YL | 7.62E+29 | -2.1±0.1 | 3.02E+29 | -1.7±0.1 |
| C/2003 K4 (LINEAR) | 20041013.7066 | -126.527 | 125.092 | 72 | 1.023580 | 2.79E-05 | DN | 4.83E+29 | -1.7±0.2 | 5.62E+29 | -1.7±0.1 |
| C/2004 Q2 (Machholz) | 20050124.9190 | -143.724 | 119.311 | 78 | 1.205709 | 4.06E-04 | YL | 5.91E+29 | -2.0±0.1 | 6.49E+29 | -3.7±0.1 |
| C/2006 M4 (SWAN) | 20060928.7285 | -79.025 | 89.000 | 51 | 0.783008 | 2.05E-04 | YL | 1.50E+29 | -2.7±0.2 | 2.35E+29 | -2.9±0.2 |
| C/2006 P1 (McNaught) | 20070112.7991 | -35.886 | 77.439 | 60 | 0.170752 | 2.98E-05 | DN | 7.40E+29 | -2.1±0.1 | 5.20E+29 | -2.5±0.1 |
| C/2007 F1 (LONEOS) | 20071028.7593 | -43.520 | 34.790 | 47 | 0.402402 | -1.70E-04 | DN | 6.11E+28 | -1.0±0.1 | 8.19E+27 | -3.9±0.1 |
| C/2007 N3 (Lulin) | 20090110.6117 | 0.261 | 101.321 | 63 | 1.211769 | 3.12E-05 | DN | b | b | 1.78E+29 | -1.7±0.1 |
| C/2009 K5 (McNaught) | 20100430.0228 | -13.775 | 27.522 | 32 | 1.422394 | 4.94E-05 | DN | b | b | 1.46E+29 | -0.7±0.5 |
| C/2009 R1 (McNaught) | 20100702.6852 | -56.764 | 47.193 | 64 | 0.405037 | -4.51E-05 | DN | 1.28E+29 | -1.9±0.1 | 6.63E+28 | -2.7±0.1 |
| C/2009 P1 (Garradd) | 20111223.6776 | -129.718 | 105.384 | 117 | 1.550537 | 4.20E-04 | YL | 2.76E+29 | -0.5±0.2 | 6.77E+29 | -3.2±0.2 |

| Comet | T-Perihelion | T-Begin | T-End | N | q(AU) | $a_0$ (AU) | Class | Q1-pre | p-pre | Q1-post | p-post |
|---|---|---|---|---|---|---|---|---|---|---|---|
| C/2012 E2 (SWAN) | 20120315.0351 | -12.084 | -6.112 | 5 | 0.007164 | 2.36E-04 | YL | 5.11E+26 | -3.3±0.4 | b | b |
| C/2011 L4 (PanSTARRS) | 20130310.1699 | -44.470 | 51.389 | 50 | 0.301545 | 2.96E-05 | DN | 1.30E+29 | -1.3±0.1 | 5.00E+28 | -2.3±0.3 |
| C/2012 F6 (Lemmon) | 20130324.5151 | -105.443 | 99.409 | 109 | 0.731248 | 2.19E-03 | OL | 3.20E+29 | -3.0±0.1 | 4.80E+29 | -2.3±0.1 |
| C/2012 S1 (ISON) | 20131128.7792 | -34.901 | -5.148 | 22 | 0.012444 | -7.57E-05 | DN | 1.08E+29 | -3.0±0.1 | b | b |
| C/2013 R1 (Lovejoy) | 20131222.7339 | -69.730 | 91.039 | 141 | 0.811819 | 2.75E-03 | OL | 7.61E+28 | -2.2±0.1 | 9.09E+28 | -1.6±0.1 |
| C/2012 K1 (PanSTARRS) | 20140827.6559 | -126.088 | 102.044 | 138 | 1.054531 | 2.46E-05 | DN | 2.00E+29 | -0.8±0.2 | 1.90E+29 | -2.4±0.3 |
| C/2013 V5 (Oukaimeden) | 20140928.2249 | -38.435 | 10.633 | 41 | 0.625506 | 8.36E-06 | DN | 3.40E+28 | -1.0±0.1 | 7.0E+28 | -0.2±0.02 |
| C/2014 Q2 (Lovejoy) | 20150130.0781 | -61.833 | 122.456 | 138 | 1.290484 | 1.99E-03 | YL | 2.30E+30 | -6.6±0.6 | 1.90E+30 | -3.4±0.1 |
| C/2014 Q1 (PanSTARRS) | 20150706.5128 | -46.609 | 34.885 | 31 | 0.314557 | 1.20E-03 | YL | 5.30E+27 | -7.8±0.1 | 9.10E+26 | -8.9±0.1 |
| C/2013 US10 (Catalina) | 20151115.7201 | -137.27 | 97.077 | 119 | 0.822956 | 5.97E-05 | DN | 2.80E+29 | -1.6±0.1 | 1.60E+29 | -1.9±0.1 |
| C/2014 E2 (Jacques) | 20140702.5164 | -88.844 | 110.226 | 97 | 0.663946 | 1.25E-03 | OL | 1.50E+29 | -2.4±0.2 | 1.10E+29 | -1.7±0.1 |
| C/2015 G2 (MASTER) | 20150523.8022 | -45.896 | 34.098 | 59 | 0.779772 | 5.45E-05 | YL | 4.10E+28 | -1.8±0.2 | 6.65E28 | -0.7±0.3 |
| C/2013 X1 (PanSTARRS) | 20160420.7226 | -127.769 | 102.222 | 121 | 1.314254 | 2.41E-04 | YL | 8.20E+29 | -3.0±0.1 | 2.70E+29 | -2.2±0.1 |

Notes to Table 1
T-Perihelion - Date of perihelion yyyymmdd.frac
T-Begin - Start time image in days from perihelion
T-End - End time image in days from perihelion
N - number of images
q(AU) - perihelion distance in AU
$a_0$ (AU) - original semi-major axis
Class - DN (dynamically new), YL (young long period), OL (old long period), following A'Hearn et al. (1995), see text for definition
Q1-pre - production rate at 1 AU from power law fitted to pre-perihelion observations
p-pre - power-law exponent fitted to pre-perihelion observations
Q1-post - production rate at 1 AU from power law fitted to post-perihelion observations
p-post - power-law exponent fitted to post-perihelion observations
a. A power law did not represent the variation of Q with r.
b. No or too little data available.

Table II. Observational and Orbital Aspects of Short Period Comets Observed by SOHO SWAN

| Comet | T-Perihelion | T-Begin | T-End | N | q | e | i | a | $T_J$ |
|---|---|---|---|---|---|---|---|---|---|
| 2P/Encke (1997) | 19970523.5987 | -3.179 | 34.408 | 15 | 0.331396 | 0.850014 | 11.9294 | 2.209513 | 3.026452 |
| 2P/Encke (2000) | 20000909.6611 | -38.75 | 34.765 | 18 | 0.339537 | 0.846898 | 11.7557 | 2.217718 | 3.025800 |
| 2P/Encke (2003) | 20031229.8768 | -28.244 | -2.477 | 14 | 0.338461 | 0.847339 | 11.7696 | 2.217076 | 3.025448 |
| 2P/Encke (2007) | 20070419.3117 | -30.174 | 53.836 | 38 | 0.339269 | 0.847040 | 11.7543 | 2.218024 | 3.025237 |
| 2P/Encke (2010) | 20100806.5019 | 7.242 | 38.163 | 7 | 0.335869 | 0.848338 | 11.7831 | 2.214589 | 3.025628 |
| 2P/Encke (2013) | 20131121.6945 | -45.701 | -4.063 | 38 | 0.336127 | 0.848232 | 11.7790 | 2.214742 | 3.025716 |
| 2P/Encke (2017) | 20170310.0924 | -42.109 | 40.938 | 57 | 0.335895 | 0.848334 | 11.7782 | 2.214702 | 3.025546 |
| 8P/Tuttle (2008) | 20080127.0256 | -18.759 | 21.493 | 33 | 1.027117 | 0.819800 | 54.9832 | 5.699872 | 1.600663 |
| 9P/Tempel 1 (2005) | 20050705.3153 | -0.577 | 24.875 | 9 | 1.506167 | 0.517491 | 10.5301 | 3.121531 | 2.970018 |
| 17P/Holmes (2007) | 20070419.3117 | 174.338 | 199.001 | 23 | 0.339269 | 0.847040 | 19.1132 | 2.218024 | 3.001528 |
| 19P/Borrelly (2001) | 20010914.7055 | -72.586 | 106.195 | 76 | 1.358096 | 0.623798 | 30.3238 | 3.610018 | 2.565162 |
| 21P/Giacobini-Zinner (1998) | 19981121.3205 | -1.208 | 28.218 | 18 | 1.033713 | 0.706483 | 31.8588 | 3.521816 | 2.466428 |
| 21P/Giacobini-Zinner (2005) | 20050702.7605 | -54.877 | 26.31 | 39 | 1.037914 | 0.705691 | 31.8109 | 3.526613 | 2.466711 |
| 21P/Giacobini-Zinner (2012) | 20120211.7347 | 45.490 | 27.158 | 37 | 1.030489 | 0.707045 | 31.9106 | 3.517568 | 2.466276 |
| 24P/Schaumasse (2001) | 20010502.6917 | 9.592 | 34.828 | 4 | 1.205203 | 0.704923 | 11.7505 | 4.084368 | 2.504373 |
| 41P/Tuttle-Giacobini-Kresak (2001) | 20010106.9664 | -47.354 | 23.318 | 34 | 1.052233 | 0.659237 | 9.2255 | 3.087873 | 2.828496 |
| 41P/Tuttle-Giacobini-Kresak (2006) | 20060611.7914 | -40.537 | 31.679 | 27 | 1.047777 | 0.660413 | 9.2280 | 3.085445 | 2.827795 |
| 41P/Tuttle-Giacobini-Kresak (2017) | 20170412.7519 | -12.081 | 8.623 | 13 | 1.045044 | 0.661243 | 9.2293 | 3.084937 | 2.826863 |
| 45P/Honda-Mrkos-Pajdusakova (2001) | 20010329.9268 | -2.952 | 50.023 | 21 | 0.528411 | 0.825078 | 4.2556 | 3.020838 | 2.581000 |
| 45P/Honda-Mrkos-Pajdusakova (2011) | 20110928.7813 | -33.093 | 46.142 | 46 | 0.529641 | 0.824626 | 4.25233 | 3.020066 | 2.582336 |
| 45P/Honda-Mrkos-Pajdusakova (2017) | 20161231.2670 | -15.322 | 47.465 | 26 | 0.532555 | 0.823995 | 4.24940 | 3.025795 | 2.581290 |
| 46P/Wirtanen (1997) | 19970314.1501 | -50.819 | 47.957 | 44 | 1.063764 | 0.656748 | 11.7225 | 3.099076 | 2.818571 |
| 46P/Wirtanen (2002) | 20020826.6791 | -43.643 | 64.142 | 28 | 1.058618 | 0.657747 | 11.7381 | 3.093086 | 2.819340 |
| 55P/Tempel-Tuttle (1998) | 19980228.0975 | -56.783 | 59.802 | 36 | 0.976576 | 0.905518 | 162.4862 | 10.336106 | 0.637301 |
| 67P/Churyumov-Gerasimenko (1996) | 19960117.6564 | 5.173 | 34.15 | 4 | 1.300034 | 0.630192 | 7.1133 | 3.515430 | 2.746628 |
| 67P/Churyumov-Gerasimenko (2002) | 20020818.2877 | -2.723 | 27.736 | 10 | 1.290650 | 0.631750 | 7.1240 | 3.504820 | 2.747102 |
| 67P/Churyumov-Gerasimenko (2009) | 20090228.2985 | 2.231 | 50.378 | 28 | 1.247284 | 0.640337 | 7.0418 | 3.467924 | 2.744986 |
| 73P/Schwassmann-Wachmann 3 (2001) | 20010127.7149 | -24.731 | 80.026 | 52 | 0.937355 | 0.693815 | 11.4061 | 3.061401 | 2.782496 |
| 73P/Schwassmann-Wachmann 3-B (2006) | 20060607.9244 | -65.189 | -9.886 | 24 | 0.939111 | 0.693286 | 11.3970 | 3.061846 | 2.783129 |
| 73P/Schwassmann-Wachmann 3-C (2006) | 20060606.9570 | -66.264 | 53.485 | 52 | 0.939146 | 0.693207 | 11.3958 | 3.061172 | 2.783503 |

| Comet | T-Perihelion | T-Begin | T-End | N | q(AU) | e | i | a | $T_J$ |
|---|---|---|---|---|---|---|---|---|---|
| 81P/Wild 2 (1997) | 19970506.6272 | -37.071 | 53.649 | 33 | 1.582622 | 0.540221 | 3.2426 | 3.442136 | 2.878274 |
| 96P/Machholz 1 (1996) | 19961015.0696 | -15.728 | 26.043 | 8 | 0.124718 | 0.958637 | 60.0742 | 3.015207 | 1.941717 |
| 96P/Machholz 1 (2002) | 20020108.6268 | -14.317 | 25.249 | 12 | 0.124110 | 0.958243 | 60.1857 | 2.972196 | 1.965408 |
| 96P/Machholz 1 (2007) | 20070404.6194 | -23.725 | 25.39 | 24 | 0.124618 | 0.958684 | 59.9553 | 3.016216 | 1.941834 |
| 96P/Machholz 1 (2012) | 20120714.7852 | 10.194 | 20.198 | 4 | 0.123791 | 0.959182 | 58.2989 | 3.032755 | 1.942448 |
| 103P/Hartley 2 (1997) | 19971222.0173 | -50.518 | 97.534 | 50 | 1.031722 | 0.700375 | 13.6189 | 3.443378 | 2.639661 |
| 103P/Hartley 2 (2010) | 20101028.2570 | -42.853 | 45.608 | 61 | 1.058690 | 0.695145 | 13.6169 | 3.472766 | 2.639781 |
| 141P/Machholz 2 (1999) | 19991209.9566 | -41.387 | 47.507 | 13 | 0.748991 | 0.751078 | 12.8119 | 3.008939 | 2.708274 |

Notes for Table 2
T-Perihelion = Time /Date of perihelion yyyymmdd.fraction
T-Begin = first image in days from perihelion
T-End = last image in days from perihelion
N = number of images
q(AU) = perihelion distance in AU
e = eccentricity of orbit
i = inclination of orbit in degrees
a = semi-major axis of orbit in AU
$T_J$ = Tisserand constant with Jupiter

Table 3. Observational Results of Individual Apparitions of Short Period Comets Observed with SOHO SWAN

| Comet | $r_N$ | Q-1AU | p-pre | Q-1AU | p-post | AApre | AApost | AFpre | AFpost |
|---|---|---|---|---|---|---|---|---|---|
| 2P/Encke (1997) | 2.40 | b | b | 2.13E+28 | -1.0±0.3 | - | 1.34 | - | 0.0185 |
| 2P/Encke (2000) | 2.40 | 8.18E+27 | -3.1±0.7 | 1.01E+28 | -3.2±0.5 | 0.51 | 0.63 | 0.0071 | 0.0088 |
| 2P/Encke (2003) | 2.40 | 9.98E+27 | -1.0±0.2 | b | b | 0.63 | - | 0.0086 | - |
| 2P/Encke (2007) | 2.40 | 1.60E+28 | -0.7±0.3 | 9.12E+27 | -2.0±0.2 | 1.00 | 0.57 | 0.0139 | 0.0079 |
| 2P/Encke (2010) | 2.40 | b | b | 5.11E+27 | -2.4±0.5 | | 0.32 | | 0.0044 |
| 2P/Encke (2013) | 2.40 | 4.58E+28 | -2.4±0.1 | b | b | 2.87 | | 0.0397 | |
| 2P/Encke (2017) | 2.40 | 7.93E+27 | -1.1±0.2 | 9.23E+27 | -2.1±0.2 | 0.50 | 0.58 | 0.0069 | 0.0080 |
| 8P/Tuttle (2008) | 2.25 | 4.11E28 | -8.9±0.1 | 3.42E28 | 3.2±0.1 | 2.58 | 2.15 | 0.0405 | 0.0337 |
| 9P/Tempel 1 (2005) | 2.72 | a | a | a | a | - | - | - | - |
| 17P/Holmes (2007) | 1.61 | a | a | b | b | - | - | - | - |
| 19P/Borrelly (2001) | 2.40 | 2.80E+29 | -5.2±1.1 | 3.50E+29 | -6.6±0.6 | 17.56 | 21.95 | 0.2426 | 0.3032 |
| 21P/Giacobini-Zinner (1998) | 1.82 | b | b | 4.04E+28 | -9.0±0.9 | - | 2.53 | - | 0.0609 |
| 21P/Giacobini-Zinner (2005) | 1.82 | 5.88E+29 | -1.7±0.4 | 8.10E+28 | -18.5±3.5 | 36.87 | 5.08 | 0.8858 | 0.1220 |
| 21P/Giacobini-Zinner (2012) | 1.82 | 2.30E+29 | -11.9±0.2 | 2.44E+29 | -11.1±0.1 | 14.42 | 15.3 | 0.3465 | 0.3268 |
| 24P/Schaumasse (2001) | b | b | b | b | b | - | - | - | - |
| 41P/Tuttle-Giacobini-Kresak (2001) | 0.70 | 2.53E+28 | --5.9±0.1 | 7.09E+29 | -7.9±0.5 | 1.59 | 44.46 | 0.2578 | 7.220 |
| 41P/Tuttle-Giacobini-Kresak (2006) | 0.70 | 1.89E+28 | -2.6±0.5 | 2.15E+28 | a | 1.19 | 1.35 | 0.1925 | 0.2190 |
| 41P/Tuttle-Giacobini-Kresak (2017) | 0.70 | 1.59E+27 | a | 4.07E+27 | a | 0.10 | 0.26 | 0.0162 | 0.0413 |
| 45P/Honda-Mrkos-Pajdusakova (2001) | 0.39 | b | b | 6.52E+27 | -3.6±0.4 | - | 0.41 | - | 0.2139 |
| 45P/Honda-Mrkos-Pajdusakova (2011) | 0.39 | 6.47E+26 | -6.6±0.1 | 1.46E+28 | -2.8±0.1 | 0.04 | 0.92 | 0.0212 | 0.4790 |
| 45P/Honda-Mrkos-Pajdusakova (2017) | 0.39 | 2.24E+27 | -4.2±0.2 | 1.27E+28 | -2.0±0.1 | 0.14 | 0.80 | 0.0735 | 0.4167 |
| 46P/Wirtanen (1997) | 0.6 | 1.78E+28 | -3.6±0.1 | 1.84E+28 | -0.9±0.1 | 1.12 | 1.15 | 0.0617 | 0.0637 |
| 46P/Wirtanen (2002) | 0.6 | 2.13E+28 | a | 4.24E+28 | -3.9±1.4 | 1.34 | 2.66 | 0.0738 | 0.1469 |
| 55P/Tempel-Tuttle (1998) | 1.80 | 4.95E+28 | -6.3±0.4 | 4.13E+28 | -1.7±0.3 | 3.10 | 2.59 | 0.0762 | 0.0636 |
| 67P/Churyumov-Gerasimenko (1996) | 2.00 | b | b | 4.43E+28 | -4.5±2.3 | - | 2.78 | - | 0.0553 |
| 67P/Churyumov-Gerasimenko (2002) | 2.00 | b | b | a | -12±11 | - | - | - | - |
| 67P/Churyumov-Gerasimenko (2009) | 2.00 | b | b | 2.32E+28 | -4.3±2.1 | - | 1.46 | - | 0.0289 |
| 73P/Schwassmann-Wachmann 3 (2001) | 1.26 | 7.40E+28 | -0.8±3.0 | 8.44E+28 | -2.1±0.3 | 4.64 | 5.29 | 0.2326 | 0.2653 |
| 73P/Schwassmann-Wachmann 3-B (2006) | b | 1.75E+28 | -1.3±1.0 | b | b | 1.10 | - | - | - |
| 73P/Schwassmann-Wachmann 3-C (2006) | b | 1.97E+28 | -2.2±0.4 | 1.23E+28 | -6.9±0.9 | 1.24 | 0.77 | - | - |

| Comet | rN | Q1-pre | p-pre | Q1-post | p-post | AA pre | AA post | AF pre | AF post |
|---|---|---|---|---|---|---|---|---|---|
| 81P/Wild 2 (1997) | 2.10 | a | a | a | a | | | | |
| 96P/Machholz 1 (1996) | 3.20 | 7.12E+26 | -3.5±0.1 | 6.84E+27 | -2.4±1.2 | 0.045 | 0.429 | 0.00035 | 0.0033 |
| 96P/Machholz 1 (2002) | 3.20 | 1.74E+27 | -2.4±0.3 | 1.46E+28 | -0.8±0.8 | 0.109 | 0.916 | 0.00085 | 0.0071 |
| 96P/Machholz 1 (2007) | 3.20 | 2.54E+27 | -2.9±0.5 | 5.04E+27 | -2.3±0.2 | 0.159 | 0.316 | 0.00124 | 0.0025 |
| 96P/Machholz 1 (2012) | 3.20 | b | b | 4.33E+27 | -2.9±0.4 | - | 0.272 | - | 0.0021 |
| 103P/Hartley 2 (1997) | 0.54 | 3.94E+28 | -6.6±0.1 | 3.08E+28 | -3.2±0.1 | 2.47 | 1.931 | 0.6743 | 0.5271 |
| 103P/Hartley 2 (2010) | 0.54 | 2.31E+28 | -14.0±1.0 | 1.35E+28 | -7.2±1.2 | 1.45 | 0.847 | 0.3953 | 0.2310 |
| 141P/Machholz 2 (1999) | b | 7.45E+27 | -0.4±0.9 | 1.79E+27 | -5.3±1.1 | 0.47 | 0.139 | - | - |

Notes for Table 3
rN = nucleus radius in km.
Q1- pre = power law water production rate in molecules s$^{-1}$ at 1 AU for pre-perihelion observations
p-pre = power law exponent for pre-perihelion observations
Q1- post = power law water production rate in molecules s$^{-1}$ at 1 AU for post-perihelion observations
p-post = power law exponent for post-perihelion observations
AA pre = active area in km$^2$ at 1 AU for pre-perihelion observations
AA post = active area in km$^2$ at 1 AU for post-perihelion observations
AF pre = active fraction at 1 AU for pre-perihelion observations
AF post = active fraction at 1 AU for post-perihelion observations
a. A power law did not represent the variation of Q with r.
b. No or too little data available.
References for nucleus radii $r_N$ in km
2P/Encke - 2.4±0.3 Fernandez et al. (2000)
8P/Tuttle - 2.25±0.5 Equivalent single sphere from Harmon et al. (2008); Harmon et al. (2010) radar observation
9P/Tempel 1 - Lamy et al. (2007) verified by s/c flyby
17P/Holmes - Snodgrass, Lowry and Fitzsimmons (2006)
19P/Borrelly - 2.40 Weaver et al. (2003), verified by s/c flyby
21P/Giacobini-Zinner - Pittichova et al. (2008)
41P/ Tuttle-Giacobini-Kresak - Lamy et al (2004); Howell et al. (2017) r>0.45 km radar
46P/Wirtanen -- Lamy et al (2004) gives individual values of 0.62 0.56, 0.7, 0.6
45P/HMP - Fernandez, J. (private communication); Lowry et al. (2003) 1.34km; Tancredi et al. 0.33 km;
67P/Churyumov-Gerasimenko -- Lamy et al. (2007), mean spherical radius agrees with Rosetta images.
46P/Wirtanen - Lamy et al. (1998)
55P/Tempel-Tuttle - Lamy et al. (2004) Comets II
73P/SW3 in 2001 - Boehnhardt et al. (1999)
81P/Wild 2 - Fernandez et al. (2013)
96P/Machholz 2 - Schleicher (2008); Lamy et al. (2004) quotes values of (3.5, 2.8 and 3.2) that gives 3.2 ±0.2
103P/Hartley 2 - Lisse et al. (2006) verified by s/c flyby

Table 4. Average Results for Multiple-Apparitions of Short Period Comets Observed with SOHO SWAN

| Comet | $r_N$ | Q1-pre | p-pre | Q1-post | p-post | AApre | AApost | AFpre | AFpost |
|---|---|---|---|---|---|---|---|---|---|
| 2P/Encke (1997, 2000, 2003, 2007, 2010, 2013, 2017) | 2.40 | 6.59E+27 | -1.8±0.1 | 1.04E+28 | -2.1±0.2 | 0.58 | 0.65 | 0.0080 | 0.0090 |
| 21P/Giacobini-Zinner (1998, 2005, 2012) | 1.82 | 4.79E+28 | -1.2±0.8 | 4.54E+28 | -11.7±1.6 | 3.00 | 2.85 | 0.0722 | 0.0684 |
| 45P/Honda-Mrkos-Pajdusakova (2000, 2011, 2017) | 0.39 | 8.55E+26 | -5.9±0.3 | 6.26E+27 | -3.7±0.2 | 0.05 | 0.39 | 0.0281 | 0.2054 |
| 46P/Wirtanen (1997, 2002) | 0.6 | 1.68E+28 | -3.6±0.7 | 2.89E+28 | -3.4±1.3 | 1.82 | 1.09 | 0.4615 | 0.2769 |
| 96P/Machholz (1996, 2002, 2007, 2012) | 3.20 | 2.52E+27 | -2.5±0.7 | 6.77E+27 | -1.9±0.2 | 0.16 | 0.42 | 0.0012 | 0.0033 |

Notes for Table 4
Comet name with perihelion year of apparitions included in the averages
rN = nucleus radius in km.
Q1- pre = power law water production rate in molecules s$^{-1}$ at 1 AU for pre-perihelion observations
p-pre = power law exponent for pre-perihelion observations
Q1- post = power law water production rate in molecules s$^{-1}$ at 1 AU for post-perihelion observations
p-post = power law exponent for post-perihelion observations
AA pre = active area in km$^2$ at 1 AU for pre-perihelion observations
AA post = active area in km$^2$ at 1 AU for post-perihelion observations
AF pre = active fraction at 1 AU for pre-perihelion observations
AF post = active fraction at 1 AU for post-perihelion observations

Table 5. Taxonomic Classes of SWAN Comets

**Typical Short Period:** 2P/Encke, 8P/Tuttle, 9P/Tempel 1, 45P/Honda-Mrkos-Pajdusakova, 46P/Wirtanen, 55P/Tempel-Tuttle, 67P/Churyumov-Gerasimenko, 103P/Hartley 2, 153P/ Ikeya-Zhang, 141P/Machholz 2

**Depleted Short Period:** 19P/Borrelly, 21P/Giacobini-Zinner, 81P/Wild 2, 96P/Machholz 1

**Typical Long Period:** C/1996 B2 (Hyakutake), C/1995 O1 (Hale-Bopp), C/1999 H1 (Lee), C/2001 A2 (LINEAR), C/2000 WM1 (LINEAR), C/2004 Q2 (Machholz), C/2007 N3 (Lulin), C/2009 R1 (McNaught), C/2012 F6 (Lemmon), C/2012 S1 (ISON), C/2014 Q2 (Lovejoy)

**Depleted Long Period:** C/1999 S4 (LINEAR), C/2012 K1 (PanSTARRS), C/2013 R1 (Lovejoy)

Table 6. Comparison of Water Production Power-law Exponents with Taxonomic Class

| Comet | Number of Comets | Power-Law Exponent |
|---|---|---|
| Typical Short Period | 4 | -3.25 |
| Depleted Short Period | 3 | -4.33 |
| Typical Long Period | 11 | -2.62 |
| Depleted Long Period | 3 | -1.82 |